# Immersed in Reality Secured by Design -A Comprehensive Analysis of Security Measures in AR/VR Environments

Authors : Sameer chauhan , Luv sachdeva (Vit Vellore University)

*Abstract-.* Virtual reality and related technologies such as mixed and augmented reality have received extensive coverage in both mainstream and fringe media outlets. When the subject goes to a new AR headset, another AR device, or AR glasses, the talk swiftly shifts to the technical and design details. Unfortunately, no one seemed to care about security. Data theft and other forms of cyberattack pose serious threats to virtual reality systems. Virtual reality goggles are just specialist versions of computers or Internet of Things devices, whereas virtual reality experiences are software packages. As a result, AR systems are just as vulnerable as any other Internet of Things (IoT) device we use on a daily basis, such as computers, tablets, and phones. Preventing and responding to common cybersecurity threats and assaults is crucial. Cybercriminals can exploit virtual reality headsets just like any other computer system. This paper analysis the data breach induced by these assaults could result in a variety of concerns, including but not limited to identity theft, the unauthorized acquisition of personal information or network credentials, damage to hardware and software, and so on. Augmented reality (AR) allows for real-time monitoring and visualization of network activity, system logs, and security alerts. This allows security professionals to immediately identify threats, monitor suspicious activities, and fix any issues that develop. This data can be displayed in an aesthetically pleasing and intuitively structured format using augmented reality interfaces, enabling for faster analysis and decision-making.

## INTRODUCTION

### A. Background

A virtual reality headset is only one component of a complicated system, and this must be carefully studied. Virtual reality headsets usually connect to a content marketplace where users can download a variety of apps. Malicious persons may launch an attack on the market, an app, or even the virtual reality gadget. In contrast, virtual reality provides cybercriminals with unprecedented chances [7]. Mark Zuckerberg of Facebook and other well-known tech CEOs are now developing metaverse technology. This technology enables us to explore AR environments by understanding our physical gestures, such as reaching, nodding, stepping, and blinking, using headgear. Gaming, socializing, business meetings, and possibly even purchasing or completing transactions can all occur in these contexts.

### B. Aim, Objectives, and Research Questions

*Aim*

This paper aims to draw on Cybersecurity in Augmented Reality (AR) and Virtual Reality (VR) Environments.

*Objectives*
- To reflect on Cybersecurity in Augmented Reality (AR) and Virtual Reality (VR) Environments.
- To establish a relationship between Cybersecurity in Augmented Reality (AR) and Virtual Reality (VR) Environments.
- To enlist the issues while maintaining data security in VR and A settings.
- To highlight the countermeasures against the issues identified.

*Research questions*
- What is the concept of Cybersecurity in Augmented Reality (AR) and Virtual Reality (VR) Environments?
- What is the relationship between Cybersecurity in Augmented Reality (AR) and Virtual Reality (VR) Environments?
- What are the issues while maintaining data security in VR and A settings?
- What are the countermeasures against the issues identified?

### C. Research Rationale

A group of computer scientists from UCR's Bourns College of Engineering, led by professors Jiasi Chen and Nael Abu-Ghazaleh, demonstrated that spyware can record and observe our every move. This spyware can also transform these motions into words with an accuracy rate of 90% or higher using artificial intelligence. This finding emphasizes the importance of personal information in the cyber domain. Any VR system must be able to track and record the user's movements as they interact with the environment [11]. However, many people are unlikely to be aware that their gait is as distinct as their fingerprint. Without the user's knowledge or permission, a firm or bad actor might possibly identify them at any time by collecting and analyzing the aforementioned movement data. Cybercriminals could utilize the user's location data to deceive them into providing personal information. Furthermore, this takes us to another distinct cyber concern specific to the AR space: identification.

## LITERATURE REVIEW

Everything in virtual reality is artificial, which means it doesn't exist in the real world. Actually, there are numerous signals that people exhibit that can tell one whether a location is safe to visit, who someone is, and what is true about a scenario [4]. Virtual reality enables believable simulation and manipulation of all components. The user may have excellent intentions, yet their experience is fabricated. Imagine an organization conducting digital transactions in a simulated bank. The money being moved is real, but how can one determine whether the bank, teller, or transaction record is legitimate? Unfortunately, making a definitive decision is not possible. There is yet to be a viable solution to this critical cybersecurity concern.

As per Alzahrani and Alfouzan (2022) while augmented reality browsers enable augmentation, third-party apps and organizations are responsible for developing and delivering the actual content. Because augmented reality (AR) is still in its early stages and methodologies for developing and providing verified material are evolving, questions about the sector's reliability arise. Hackers with advanced skills can deceive individuals or spread misleading information by replacing the user's augmented reality (AR) with theirs [1].

According to Alqahtani and Kavakli-Thorne, 2020), the worldwide AR market will be valued $26.9 billion by 2027. Furthermore, approximately 15% of American adults currently use virtual reality. Because of the high number of users (110.1 million in the United States alone) and augmented reality (AR) (over 171 million worldwide), there are growing concerns about AR's cybersecurity risks. In 2022, approximately 25% of virtual reality gaming consumers will be aged 25 to 34, adding to the industry's revenue of $12.13 billion. When used with biometric authentication, augmented reality overlays can significantly improve access control systems [2]. Augmented reality devices can capture biometric data, such as face recognition, which can be used to confirm the identity of persons attempting to access restricted areas or systems. This technique reduces the possibility of unauthorized access. New concerns regarding privacy have evolved as a result of the vastness, volume, and delicate nature of the data that AR devices may capture.

System designs, threat landscapes, and network topologies are all examples of complex security data that could benefit from augmented reality (AR) visualization [14]. As a result, security analysts will have an easier time identifying trends, correlations, and potential flaws, resulting in improved security measures and reduced risk [13].

Regardless of the credibility of the source, numerous cyber-attacks could jeopardise the content's reliability. This category includes data manipulation, spoofing, and sniffer. Furthermore, influencing an augmented reality user's experience of reality is just one example of how social manipulation and social engineering can occur in virtual and augmented reality. Physical security measures can be enhanced by augmented reality (AR), which provides security personnel with real-time information and precise instructions [10]. Heads-up displays or augmented reality glasses can detect potential security breaches, identify authorized persons, and provide instructions for resolving security issues, providing a timely and effective response. Because of the inherent unpredictable nature of information, augmented reality technologies can effectively manipulate humans in order to carry out social engineering attacks. One example is how hackers might persuade people to act in ways that benefit them by creating phoney signs or displays that distort their perception of reality [1].

Cybersecurity and Augmented Reality (AR) have been around for a long, but in recent years, they have grown and improved dramatically. These technologies have several possible applications, although they are most typically encountered in the business, industrial, tourist, academic, and social realms (including gaming, entertainment, and communication industries). Everyone agrees that these technologies are critical to the global Industry 4.0 project, which aims to ignite a fourth industrial revolution [5]. Not only that, but they are among the most significant technological advances of the past century. There are numerous businesses that employ Augmented Reality (AR), but one of the most significant is providing clients with critical information in a virtual environment. Despite this, numerous cybersecurity concerns have evolved as a result of ICT breakthroughs; to address these, we require comprehensive evidence-based approaches to cybersecurity policy, architecture, design, and technology [5]. Several articles provide detailed studies of the various applications of augmented reality and cybersecurity technologies, proving its promise in a variety of sectors. A lack of literature outlining the various applications of smart cities is also a challenge. Importantly, no academic work has yet combined cybersecurity studies with augmented reality (AR) in this specific context. The new Metaverse region is generating a lot of attention and excitement since it incorporates Augmented Reality (AR) and Virtual Reality (VR), both of which were previously regarded to be science fiction concepts. Privacy considerations exist, as are the normal issues associated with connecting AR and VR devices. As such, they are part of the rapidly expanding worldwide network known as the Internet of Things (IoT). Furthermore, just like any other piece of networked gear, they must monitor the flow and storage of data, which hackers may steal [9]. There are legitimate concerns about the immersive experiences that AR and VR may give, particularly in terms of social engineering and corporate espionage. When technology advances to the point that it can trick the brain into believing something is different, there will be plenty of opportunities to exploit people's vulnerabilities [17].

Users are typically the weakest link in technology systems, making AR and VR ideal targets for misuse. Criminals may, for example, insert features into VR systems to deceive users into disclosing critical information. Furthermore, ransomware has new concerns, such as the risk that attackers could intentionally harm platforms and disrupt critical meetings before demanding payment [18].

Virtual and augmented reality experiences will become increasingly lifelike and immersive. Hopefully, this will boost their involvement and utilization. On the other hand, it heightens the risk they confront. Machine learning technologies enable the creation of synthetic identities, often known as "deepfakes," which allow voice and video to be manipulated to appear authentic. If a hacker obtains motion-tracking data from a VR headset, they might theoretically make a digital duplicate. Then, to launch a social engineering attack, they can overlay this on top of another person's VR experience, such as an immersive corporate meeting. Another concerning aspect of cybercrime is the growing use of avatars as a form of interpersonal communication [6].

*Literature Gap*

The paper in consideration concentrates on professionals who are currently examining the potential ramifications of AR technologies. The use of augmented reality suggested through the current literature demonstrates that technologies will undoubtedly contribute to the growth of the cybersecurity domain, given the possibilities and demonstrated benefits of such tools in other industries [7].

There is a lack of studies regarding AR issues malicious actors can record and view people's behaviors and interactions in an augmented reality environment. Their ownership over this sensitive data enables them to hold it for ransom by generating fear of public disclosure. One major concern with augmented reality is privacy. The user's privacy is jeopardized since augmented reality technologies can observe their actions. Based on a critical examination of the presented literature a lack of study insights is seen on the notion that Augmented reality (AR) collects more data from consumers than social media and other online platforms combined. Another security worry with augmented reality devices is the possibility of physical damage or theft.

The concern for one's privacy rates among the top worries and the present scholar studies fail to address this, therefore making it a gap to fill. In comparison to other technologies and social media platforms, augmented reality (AR) apps frequently capture far more data. This includes information on the user's identity as well as their present behaviours. It raises numerous questions and concerns concerning user privacy, data storage, hacking, and its application and security. Spoofing, sniffers, and data manipulation are prominent cyber threats associated with augmented reality content. This diminishes the trustworthiness of the content, regardless of its source. Reliable methods of creating and delivering content for augmented reality technology are still in the works. Augmented reality, as a tool for social engineering, has the potential to fool users.

## METHODOLOGY

*A. Used tools and techniques*

The research focused on the positive features of integrating cybersecurity into AR and VR settings. This viewpoint holds that knowledge emerges from actual, quantitative experiences. In this context, a positivist attitude is evidence-based and takes a systematic approach [8]. Qualitative measures such as system efficiency, integration success rate, and error rate influence the problem's resolution.

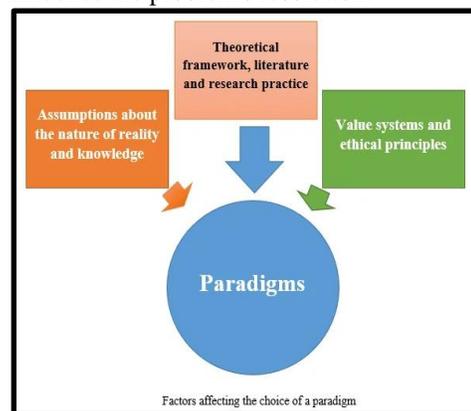

**Figure 1: Positivism philosophy [Source: Park, et. al., 2020]**

The inquiry was successful because it utilized thematic and secondary qualitative analysis. Rather of collecting new data, this allowed for a thorough examination of existing data sources. Databases, academic publications, case studies, and company reports are just a few of the sources that have been meticulously reviewed. This study drew on previous research and evaluations to acquire thorough data [9]. The utilization of secondary data collection approaches was beneficial, and the theme approach helped to reveal themes and patterns in the qualitative data.

The existing present research on the topic revolves primarily around the use of AR and VR tools. This data is implicative for this study to identify patterns, correlations, and hypotheses, then evaluated through a cybersecurity perspective. Using a top-down paradigm based on qualitative evidence, the study both confirmed prior ideas and supplied a new perspective.

*B. Data collection*

The data for this study came from secondary sources, including scholarly journals, company reports, case studies, and white papers. These publications provided quantitative outcomes for the integration of cybersecurity into AR and VR situations. The study focuses on items produced in recent years, using databases and digital libraries to verify the relevance of the content. To undertake a complete study on the subject, data was painstakingly collected, classified, and prepared for qualitative thematic analysis.

*C. Data analysis*

We conducted a thorough qualitative thematic analysis after collecting the data. Statistical methods were used for data analysis and visualization. Historically, descriptive statistics have proven useful in exposing latent trends and patterns. Inferential methods were utilized to identify relationships and evaluate literature-based assumptions. The qualitative data has been classified into numerous themes. Each individual underwent a detailed analysis to discover patterns, outliers, and recurring concerns [11].

**Table 2.1 Shows Analysis of Common AR and VR Threats: Threat Percentage Distribution**

| Common threats in AR & VR | PERCENTAGE |
|---|---|
| Latency problem | 8% |
| Observation attacks for graphical pin in 2D | 15% |
| Confidentiality, Integrity, Availability (CIA) | 15% |
| Privacy concerns | 15% |
| Data Theft | 16% |
| Security, privacy and safety (SPS) | 31% |

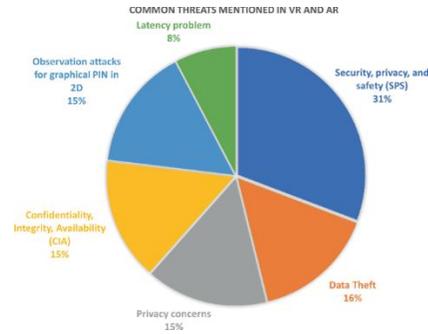

**Figure 2: Common AR and VR Threats pie chart [Source: Beitzel et al., 2018]**

By comparing these themes, it was concluded that they were appropriate and aligned with the study's purpose. This conclusion was reached based on the pie chart that was presented earlier, which depicts frequent threats posed by augmented reality and virtual reality. With the help of the stringent methods that were implemented, the team was able to extract important insights from the massive information.

*D. Ethical consideration*

Throughout the paper we prioritized ethical integrity. To limit the likelihood of plagiarism allegations while using secondary data, the study ensured that all major sources were appropriately credited. We were certain not to use any non-public or unapproved data in order to keep the material confidential [12]. There was no manipulation of the data utilized in the study to make it more compatible with the narrative. To ensure neutrality, we have disclosed any potential conflicts of interest that may have influenced our study.

FINDINGS AND ANALYSIS

Even while cyber-security and AR have both been around for a while, the development and advancement of these areas in recent years has been nothing short of remarkable. Among the many potential domains where these developments could find use are the business, industrial, tourism, academic, and social spheres. Furthermore, these advancements have the potential to positively impact other sectors, including gaming, entertainment, and communication. Industry 4.0 is a worldwide movement with the goal of launching a fourth industrial revolution; all sides agree that these technologies are crucial to this movement. They also rank highly among the most important innovations of this century, being among the most groundbreaking technological achievements. One of the most notable features of augmented reality is its capacity to provide users with fundamental details about virtual processes and procedures.

**Table 3.1- Showing Risk Mitigation Strategies for AR and VR: Number of Strategies Implemented**

| Risk mitigation strategies for AR and VR | No of Techniques used |
|---|---|
| Development of an authentication system | 4 |
| Development of secure and safe AR applications | 3 |
| Implementation of code of ethics during AR applications development | 5 |
| Utilization manipulable 3D objects for frequent authentication in VR | 4 |
| Implementation of risk assessment approach | 5 |
| Utilization level of detail (LOD) management | 4 |

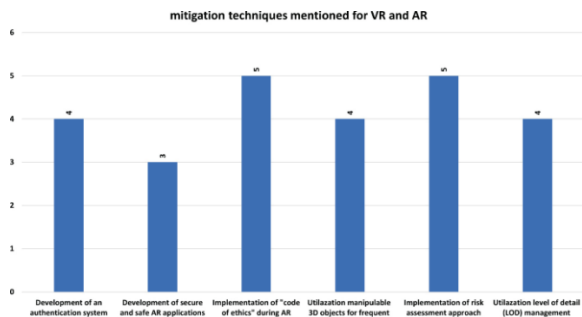

**Figure 3: Risk mitigation strategies for AR and VR [Source: Alzahrani and Alfouzan, 2022]**

As shown in the bar graph above, risk reduction measures for augmented reality and virtual reality are connected with It is totally feasible to create a virtual experience that causes a person to feel unpleasant and queasy in their stomach with the use of virtual reality. Someone with nefarious intentions could theoretically create a cyber hazard by altering a section of a virtual reality experience or adding a hidden function to make the user sick [19]. There is a possibility that someone could utilize augmented reality (AR) to intentionally distract or mislead users about their surroundings. This is despite the fact that users of AR are less likely to feel motion sickness. Virtual reality (VR) and augmented reality (AR) provide a number of distinct security challenges, some of which are the usual suspects, such as the potential for bodily injury, the leaking of sensitive information, and the malfunctioning of electrical devices.

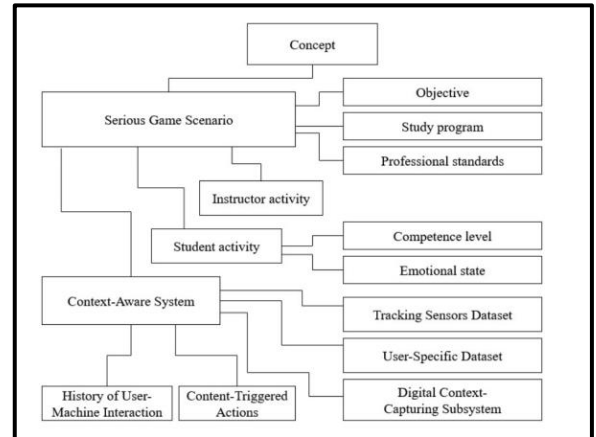

**Figure 4: Cybersecurity in Augmented reality model [Source: Skorenkyy, et. al., 2021]**

Using augmented reality (AR), cybersecurity pros may construct training simulations that are remarkably similar to the real thing. By simulating cyber-attacks in realistic situations, trainees can receive hands-on experience while honing their threat identification and mitigation abilities. This increases their abilities, discernment, and readiness to deal with real-world cyber security issues. Cameras are key components of augmented reality headsets because they allow users to perceive their immediate surroundings, but they also pose security problems. Concerns about privacy and security have been at the forefront since the 2013 launch of the Google Glass prototype, which sparked a lot of attention before being discarded by 2015 [15]. The term "glasshole," which refers to those who act badly while wearing Google Glass (for example, capturing images of people without their knowledge or consent), also appeared with the device's release. Leaving aside weird acronyms, this has significant ramifications for businesses such as healthcare, where firms are compelled by law to preserve patients' privacy (HIPAA, for example).

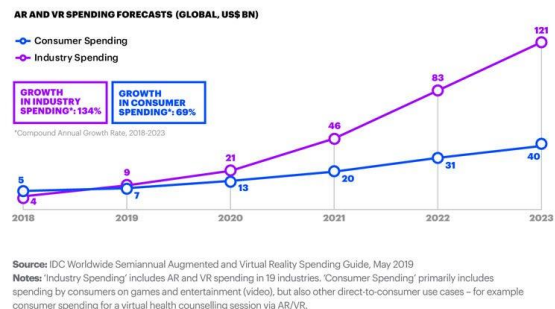

**Figure 5: AR and VR spending [Source: Odeleye, et al., 2023]**

Taking into consideration the aforementioned number concerning spending on augmented reality and virtual reality, it is clear that this goes well beyond the usual responsibilities of security experts and the cybersecurity industry. Operating on operating system devices does not protect augmented reality and virtual reality applications from the defects that are common in software. When these vulnerabilities are exploited, it is possible to take control of the device, change the data, and disrupt the user experience [12].

A data breach could occur if there is a shortage of cybersecurity specialists and staff with the necessary training. Within this paradigm, businesses seek effective training ways to create awareness and educate their employees about cybersecurity. Augmented and virtual reality technologies, when used in conjunction with gamified learning and training platforms, can help to solve this challenge. Students can interact with these AR and VR-based teaching resources in a dynamic and realistic learning environment.

To hone the abilities needed to defend against a complex cyberattack on a massive cyber-physical system. This will pave the way for systems for attack prediction, reaction creation, and strategic behaviour modelling by malicious actors. To make sure that students of cybersecurity learn what they need to know in their classes, the authors came up with the scenario model. There are substantial elements of serious gaming scenarios included in this model as well. The second part of the sentence describes how the context-aware system in the course of study interacts with the instructor, the student, and the programme itself. These exchanges revolve around the program's modules and their respective goals. The management of sensor data and data collected through user-machine interfaces is the responsibility of the context-aware system. The development of a context-aware system is underway with the goals of improving training activity control through improved user experience, management of interaction histories, and handling of user-specific data. Data acquired from different sensors, the history of human-computer interactions, and the knowledge base of the learning management system are all used to monitor and analyze all participants in an activity, like a cybersecurity incident simulation [13].

## CONCLUSION

In conclusion, some threats only materialize when lines between the real and virtual worlds start to blur. Although the risk of theft from a self-driving car is more obvious, the risk of actual bodily injury from a VR experience is substantial. Virtual reality (VR) has been reported to cause motion sickness in certain people, particularly those who use older VR headset types. The emergence of improved hardware has mitigated this. Augmented reality (AR), in particular, introduces significant hazards to users' physical security. The goal of augmented reality (AR) risk management is to identify, assess, and reduce potential hazards connected with lending money to clients and collecting past-due payments. When reviewing existing or new data protection regulations, politicians should take into account the extent to which consumers may have significant rights and firms may be held to explicit obligations regarding XR data.

To keep users, bystanders, and everyone else informed, device manufacturers should prioritize transparency in how they gather, utilize, and distribute XR data. Accounts receivable risk management is to reduce the chance of late payments and their impact on liquidity.